\documentstyle[12pt,epsf]{article}

\begin{document}
\begin{center}

{\Large {\bf
DYNAMICAL COSMOLOGICAL CONSTANT AND RELATIONS AMONG PSEUDO GOLDSTONE BOSONS} 
}
\vskip 0.5cm
{\large Juan Estrada Vigil\footnote{E-mail: estradav@ib.cnea.edu.ar} and Luis Masperi\footnote{E-mail: masperi@cab.cnea.edu.ar}}
\newline   
{\small {\it Centro At\'{o}mico Bariloche and Instituto Balseiro},}
{\small {\it (Comisi\'{o}n Nacional de Energ\'{\i}a At\'{o}mica and}}
{\small {\it Universidad Nacional de Cuyo) 8400 S.C. de Bariloche, Argentina}}

\end{center}

\vskip 0.5cm

\begin{abstract}
The present apparent cosmological constant is interpreted as the potential
of the lightest pseudo Goldstone boson. Some numerical relations among
cosmological parameters and particle masses are shown to arise from the
mixture of this very light particle, whose interaction is of the gravitational
order, with other pseudo Goldstone bosons like the pion and the axion which
feel the strong interaction. \newline
\newline
\end{abstract}

Evidence from clusters and superclusters of galaxies together with baryon
abundance indicates that matter accounts for a fraction $\Omega _m\sim 0.3$
of the critical density of the universe\cite{R1}. For the formation of large
scale structure, perhaps the age of the universe, the luminosity of galaxies
and even the reionization of intergalactic medium\cite{[2]} it is useful to
add a contribution of the cosmological constant $\Omega _\Lambda\sim 0.7$
with the formal advantage of completing the critical density required by the
standard version of inflation. However, for a correct normalization with the
COBE anisotropies it is more convenient\cite{[3]} that the apparent
cosmological constant is caused by a field which is becoming dynamical near
the present time due to the extremely light mass of the associated particle.

Cosmological parameters and typical particle masses are related by the
Friedmann equation, that does not involve $\hbar $, which determines the
expansion of the universe; and also by an intringuing relation\cite{[4]}
where $\hbar $ appears, whose validity at any time would imply the variation
of fundamental constants\cite{[5]}. This latter relation has been derived
recently\cite{[6]} to calculate the Planck constant from classical
considerations.

We show that with the dynamical cosmological constant interpretation, the
Friedmann equation takes the form of the mass of a pseudo Goldstone boson
produced by an explicit symmetry breaking term at the scale of a possible
neutrino mass, and the second relation is consistent with the hypothesis
that this very light pseudo Goldstone boson feels an interaction of the
gravitational order. The fact that the latter relation includes $\hbar $ is
due, in our interpretation, to the present situation of balance between the
microphysics term of the ligthest pseudo Goldstone boson and the
gravitational one corresponding to universe expansion.

The Hubble constant $H$ is related to the critical density $\rho _c$ and the
Newton constant $G_N$ by the Friedmann equation 
\begin{equation}
H^2=\frac{8\pi }3G_N \ \rho _c \quad ,  \label{ec1}
\end{equation}
which comes from the Einstein equation and therefore has no quantum content.
The variation of $\rho _c$ with the universe scale determines the expansion
rate through $H$.

On the other hand it has been noted that the following numerical relation
holds 
\begin{equation}
H_0=G_N \ m_\pi ^3  \label{ec2}
\end{equation}
between the present value of the Hubble constant $H_0$ and the pion mass.
This equation is written in natural units $\hbar =c=1$ but obviously
involves quantum physics by making explicit a factor $c/\hbar ^2$ on the
right hand side. If one wishes that Eq.(\ref{ec2}) holds for any time, since
it contains the Hubble constant as the only cosmological parameter some
fundamental constant should vary with time, e.g. the Newton one. A relation
analogous to Eq.(\ref{ec2}) has been obtained\cite{[6]} calculating the
Planck constant through the fluctuations of classical particles due to the
background gravitational field.

If we adopt the dynamical cosmological constant approach, an extremely light
particle should have a mass 
\begin{equation}
m_\varphi \sim H_0 \quad,  \label{ec3}
\end{equation}
so that for previous times the Hubble constant would prevail and the
contribution to energy of the field $\varphi$ would appear only through a
potential term. $\varphi$ should be therefore the lightest conceivable
matter particle in our universe.

Applying Eq.(\ref{ec1}) to present time we would have 
\begin{equation}
m_\varphi ^2\sim \frac 1{m_{Pl}^2}M^4  \label{ec4}
\end{equation}
where, due to the value of critical density, $M\sim10^{-2}-10^{-3}eV$. Eq.(\ref{ec4}) suggests that $\varphi$ is the Goldstone boson coming from a spontaneous symmetry breakdown at the Planck mass scale and which acquires a small mass through an explicit breaking of this symmetry at scale $M$. Since the numerical value of $M$ may correspond to the mass of neutrinos which solves the solar problem, it has been thought\cite{[7]} that the symmetry spontaneously broken at scale $f_\varphi \sim m_{Pl}$ is a chiral global one involving neutrinos.

The explicit breaking of this symmetry would produce an effective potential 
\begin{equation}
V(\varphi )=M^4\left( 1-\cos \frac \varphi {f_\varphi }\right)\quad,\label{ec5}
\end{equation}
due to the phase nature of $\varphi$, consistent with Eq.(\ref{ec4}).

Regarding Eq.(\ref{ec2}), it should be interpreted as
\begin{equation}
m_\varphi\sim\frac{1}{m_{Pl}^2}m_\pi ^3\quad,\label{ec6}
\end{equation}
which is satisfied for $m_\varphi\sim10^{-32}eV$ that corresponds to Eq.(\ref{ec3}). Now there is no time dependence and no need to think in changes of fundamental constants along the universe evolution, for which bounds exist\cite{[8]}.
 
The explanation of Eq.(\ref{ec6}), that must be obviously consistent with
Eq.(\ref{ec4}), may be analogous to the one which suggests the mass of the
axion. This well studied\cite{[9]} light particle is the Goldstone boson of
the spontaneous breaking of the Peccei-Quinn chiral symmetry of a quark which
consequently becomes very heavy. Since the current correponding to this
symmetry has the QCD anomaly

\begin{equation}
\partial _\mu j_{PQ}^\mu =\frac{g_S^2}{16\pi ^2}G_{\mu \nu}\widetilde{G}^{\mu \nu} 
\label{ec6_1}
\end{equation}
in terms of the gluon field $G_{\mu \nu}$, one has to subtract, apart from a factor, the axial current of light quarks to define a current which is conserved in the chiral limit

\begin{equation}
j_a^\mu =j_{PQ}^\mu -(\overline{u}\gamma ^\mu \gamma _5u+\overline{d}\gamma
^\mu \gamma _5d+\overline{s}\gamma ^\mu \gamma _5s) \quad.  \label{ec6_2}
\end{equation}

Therefore, this current will have a divergence structure analogous to that of the isospin axial current related to the symmetry whose spontaneous breaking at scale $f_\pi\sim 93MeV$ gives the pion as Goldstone boson. Due to the explicit  breaking of symmetry caused by the mass of light quarks, current algebra methods \cite{[10]} allow to calculate the mass of pion and axion in terms of  a typical quark $q$

\begin{equation}
m_\pi ^2\sim \frac{m_q}{f_\pi ^2}\langle 0|\overline{q}q|0\rangle \quad,\quad m_a^2\sim\frac{m_q}{f_a^2}\langle 0|\overline{q}q|0\rangle \label{ec6_3}
\end{equation}
apart from factors. Eq.(\ref{ec6_3}) gives the relation 
\begin{equation}
m_a\sim \frac{f_a}{f_\pi }m_\pi  \quad . \label{ec6_4}
\end{equation}

This mass is equivalent to the one coming from the effective potential V($a$%
) originated by QCD which is of the type of Eq.(\ref{ec5}) with a scale for
the explicit breaking of Peccei-Quinn symmetry $\Lambda _{QCD}$ replacing $M$.
Eq.(\ref{ec6_4}) gives the possible experimental window for axion mass $%
m_a\sim 10^{-5}eV$ if $f_a\sim10^{12}GeV$.

For the Frieman et al. chiral symmetry, we may imagine that it applies to a neutrino which becomes very heavy through spontaneous breaking at the scale $f_\varphi\sim m_{Pl}$ and that its current $j_F^\mu$ has an anomaly due to fields corresponding to interactions of gravitational order. We again define the current associated to particle $\varphi$ subtracting, apart from a factor, the axial current for a typical light neutrino to compensate this anomaly
\begin{equation}
j_\varphi ^\mu =j_F^\mu -\overline{\nu }\gamma ^\mu \gamma _5\nu 
\label{ec6_5}
\end{equation}
which will be conserved in the limit of massless
neutrino. Again current algebra methods will give 

\begin{equation}
m_\varphi ^2 \sim \frac{m_\nu}{f_\varphi^2 }\langle 0|\overline{\nu}\nu|0\rangle\quad ,
\label{ec6_5_1}
\end{equation}
and the comparison with Eq.(\ref{ec6_3}) will require

\begin{equation}
\frac{m_\nu \langle 0|\overline{\nu }\nu |0\rangle }{m_q\langle 0|\overline{q%
}q|0\rangle }\sim \left( \frac{m_\pi }{m_{Pl}}\right) ^2  \label{ec6_6}
\end{equation}
to satisfy the relation (\ref{ec6}). It is interesting to see that Eq.(\ref{ec6_6})
holds if, together with the QCD values $m_q\sim 0.1GeV$ and $\langle 0|%
\overline{q}q|0\rangle \sim 10^{-3}GeV^3,$ the estimation for the light
neutrinos scale $m_\nu \sim 10^{-2}eV$ and $\langle 0|\overline{\nu }\nu
|0\rangle \sim 10^{-6}eV^3$ is introduced.

\begin{figure}[tbp]
\hspace{2cm} \epsfysize=9cm \epsffile{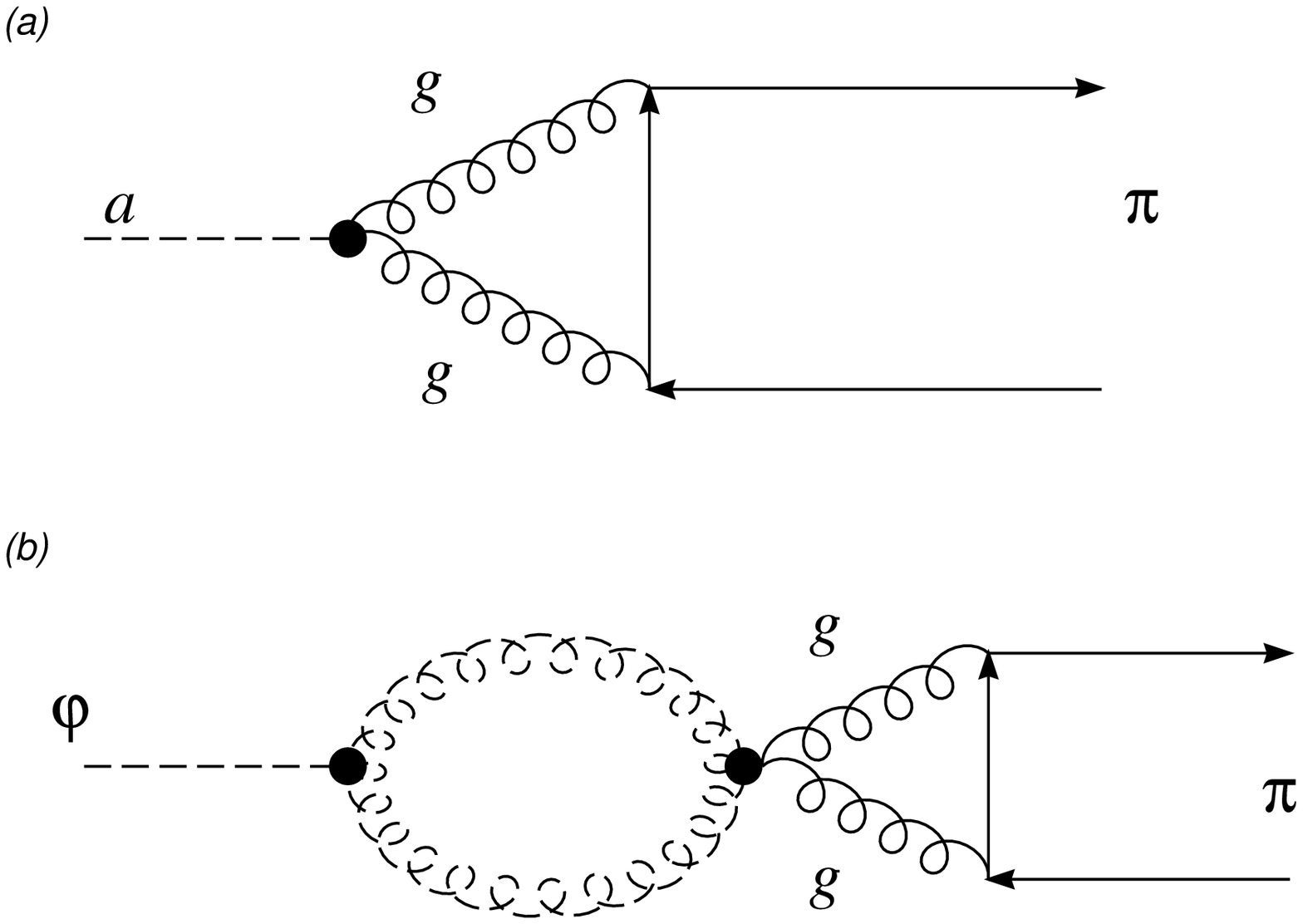}
\caption{Mixture of pseudo Goldstone bosons \protect\newline
(a) axion with
pion \protect\newline
(b) $\varphi$ with pion}
\end{figure}

The previous result may be understood also in a pictorial way. The axion has an effective coupling with two gluons through a triangle of heavy quarks, of the order of the inverse of the scale $f_a$, which corresponds to the divergence of the Peccei-Quinn current. According to Fig.1(a), since the pion is coupled to the axial current through $f_\pi$, the axion lives a fraction of its time as a pion getting a mass as given by Eq.(\ref{ec6_4}).

For the very light particle $\varphi $, due to the fact that its effective
coupling is caused by the coupling with heavy neutrinos resulting from the
spontaneous breakdown of chiral symmetry at scale $f_\varphi $, the
interaction may be thought as of gravitational order and the amplitude for the mixture with
the pion according to Fig.1(b) needs an additional conversion factor. It
seems reasonable that this factor must be the ratio of scales for strong and
gravitational interactions i.e. of the order of the ratio of the Planck and
Fermi lengths, or $m_\pi /m_{Pl}.$ Therefore the mass of the $\varphi $
particle will be 
\begin{equation}
m_\varphi \sim \frac 1{f_\varphi }\frac{m_\pi }{m_{Pl}}f_\pi m_\pi \simeq
\frac 1{m_{Pl}^2}m_\pi ^3\quad ,  \label{ec8}
\end{equation}
which is again Eq.(\ref{ec6}). 

The consistency of Eqs.(\ref{ec4}) and (\ref{ec6}) requires that the scale
for the explicit breaking of the symmetry of which $\varphi $ is the
Goldstone boson, that should be of the order of the neutrino mass, is 

\begin{equation}
M\sim \left( \frac{m_\pi }{m_{Pl}}\right) ^{1/2}m_\pi \quad .  \label{ec9}
\end{equation}

It is interesting to note that the derivation of Eq.(\ref{ec2}) done in
Ref.[6] starting from Friedmann Eq.(\ref{ec1}) together with the expression
of the energy contained in the present Hubble radius as given by N
nonrelativistic particles of mass of the order of $m_\pi ,$ is expressed in
our dynamical cosmological constant approach as the condition 
\begin{equation}
m_\pi \sim \sqrt{N}\ m_\varphi \quad.  \label{ec10}
\end{equation}

One might envisage the connection between classical and quantum mechanics of
Ref.[6] by the fact that Eq.(\ref{ec3}) indicates that the Compton length of
the $\varphi$ particle equals the radius of the observable universe. Only
from this moment, i.e. around the present time, quantum mechanics may be
applied to all particles.

Obviously the need of the variation with time of fundamental constants
assuming the general validity of Eq.(\ref{ec2}) is replaced in our scheme by
the requirement that the field $\varphi $ becomes dynamical at present which
is a form of anthropic principle \cite{[11]} even though the true
cosmological constant may be zero because of some symmetry, perhaps
supergravity\cite{[12]}. Since our relations are valid only as an order of magnitude,
 $\varphi$ might have become dynamical for a redshift $z$ of some units and 
not necesarily for $z=0$ and it is not relevant the precise value of $H_0$.

\vskip 1.0cm
{\bf Acknowledgments}

\vskip 0.5cm

We thank heartily Francesco Calogero for pointing us his
derivation of Planck constant from gravitational effects and for stimulating
discussions, and Carlos Garc\'{i}a Canal for extremely useful comments.

\end{document}